\documentclass[aps,prl,twocolumn,superscriptaddress,showpacs]{revtex4}
\usepackage{CJK}
\usepackage{graphicx}
\begin{document}
%\begin{CJK*}{GBK}{}
\title{Bound on genuine multipartite correlations from the principle of information causality}

\author{Yang Xiang}
\email{njuxy@sohu.com}
\affiliation{School of Physics and Electronics, Henan University, Kaifeng, Henan 475004, China }

\author{Wei Ren}
\email{weiren@uark.edu}
\affiliation{Physics Department, University of Arkansas, Fayetteville, Arkansas 72701, USA}
\date{\today}
\begin{abstract}
Quantum mechanics is not the unique no-signaling theory which is endowed with stronger-than-classical correlations, and there exists
a broad class of no-signaling theories allowing even stronger-than-quantum correlations.
The principle of information causality has been suggested to distinguish quantum theory from these nonphysical theories, together with an elegant information-theoretic proof of the quantum bound of two-particle correlations. In this work, we
extend this to genuine $N$-particle correlations that cannot be reduced to mixtures of states
in which a smaller number of particles are entangled.
We first express Svetlichny's inequality in terms of multipartite no-signaling boxes, then
prove that the strongest genuine multipartite correlations lead to the maximal violation of information causality. The maximal genuine multipartite correlations under the constraint of information causality is found to be equal to the quantum mechanical
bound. This result consolidates information causality as a physical principle defining the possible correlations allowed by nature,
and provides intriguing insights into the limits of genuine multipartite correlations in quantum theory.

\end{abstract}

\maketitle
%\end{CJK*}

% insert suggested keywords - APS authors don't need to do this
%\keywords{}

%\maketitle must follow title, authors, abstract, \pacs, and \keywords

% body of paper here - Use proper section commands
% References should be done using the \cite, \ref, and \label commands

{\it Introduction}~~The violation of Bell inequalities \cite{bell,chsh} proves that the quantum mechanics cannot be regarded as a local realistic theory.
Tsirelson \cite{tsir} proved an upper bound on the violation of the CHSH inequality \cite{chsh}, which means that the
amount of non-locality allowed by quantum
mechanics is limited. One may think that Tsirelson bound is a consequence
of relativity, but Popescu and Rohrlich \cite{pr} showed that there exists a broad class of no-signaling theories
which allow even stronger-than-quantum correlations. An example of the no-signaling
theories is the Popescu and Rohrlich boxes (PR-boxes) \cite{pr}. This broad class of no-signaling
theories possessing extremely powerful correlations are usually called post-quantum
theories and modeled as no-signaling boxes (NS-boxes) \cite{barrett1}.
These post-quantum theories have much in common with quantum mechanics, such
as no-cloning \cite{noclon}, no-broadcasting \cite{nobroad}, monogamy of
correlations \cite{noclon}, information-disturbance trade-offs \cite{trade}, and the security for
key distribution \cite{key}, so there is a need to find some principles at the very root of quantum theory
and distinguish it
from these post-quantum theories. In recent years, an intensive study
has been made on this issue. In Ref. \cite{dam}, van Dam showed that the availability
of PR-boxes makes communication complexity trivial. However, communication
complexity is not trivial in quantum physics and it is strongly believed
that devices producing such correlations making communication complexity trivial are very unlikely to exist.
Later, Brassard \textit{et al.} proved that some post-quantum theories would lead
to an implausible simplification of distributed computational tasks \cite{brassard1,nonlocal comput,brun1}.
More recently, Barnum \textit{et al.} \cite{barnum,acin} showed that the combination of local quantum measurement assumption and relativity
results in quantum correlations, and in Ref. \cite{acin} the authors provided a unified framework for all no-signaling theories.

From an information theoretic point of view, Paw{\l}owski \emph{et al.} \cite{pawlowski1} suggested a bold
physical principle: information causality (IC), stating that communication of $n$ classical bits
causes information gain of at most $n$ bits. When $n=0$ IC is just the no-signaling principle.
In a bipartite scenario where each party has two inputs and two outputs,
Paw{\l}owski \emph{et al.} showed that IC is respected both in classical and quantum physics, but all correlations
stronger than the strongest quantum correlations (Tsirelson bound) violate it, and they derived Tsirelson bound from IC.
It must be noted that there are some stronger-than-quantum correlations which are not known to violate IC \cite{allcock, cav}.

The present work is to extend the research of understanding the quantum mechanical bound on nonlocal correlations to
genuine multipartite correlations. The structure of multipartite correlations is much richer than that of bipartite
correlations \cite{barrett1}. For example, in \cite{pironio} the authors dealt with a tripartite scenario where each party has two inputs
and two outputs, they found that there exist $53856$ extremal no-signaling tripartite correlations which belong to $46$ inequivalent
classes, and there are more than three classes which feature genuine tripartite nonlocality. So there exist many inequivalent types of genuine
multipartite correlations, and in the present paper we deal only with Svetlichny genuine multipartite correlations which is relevant
to Svetlichny's inequality (SI) \cite{svet,seev}.  We first express SI in terms of multipartite no-signaling boxes,
and then prove that the strongest Svetlichny genuine multipartite correlation leads to the maximal violation of IC.
Under the constraint of IC, the maximal Svetlichny genuine multipartite correlation just equals to the quantum mechanical
bound.

{\it Tripartite Svetlichny's inequality}~~We first introduce SI of three-particle \cite{svet}, which can distinguish
between genuine three-particle correlations and
two-particle correlations. A violation of SI implies the presence of genuine three-particle correlations. Consider three observers, Alice, Bob, and Carol, who share three entangled qubits. Each of the three observers can choose to measure one
of two dichotomous observables. We denote $x\in\{0,1\}$ and $A\in\{-1,1\}$ as Alice's measurement choice and outcome respectively, and
similarly $y$ and $B$ ($z$ and $C$) for Bob's (Carol's). Thus SI can be expressed as \cite{svet}
%\begin{widetext}
\begin{eqnarray}
S&\equiv&|E(ABC|x=0,y=0,z=0)\nonumber\\
&&+E(ABC|x=0,y=0,z=1)\nonumber\\
&&+E(ABC|x=0,y=1,z=0)\nonumber\\
&&+E(ABC|x=1,y=0,z=0)\nonumber\\
&&-E(ABC|x=0,y=1,z=1)\nonumber\\
&&-E(ABC|x=1,y=0,z=1)\nonumber\\
&&-E(ABC|x=1,y=1,z=0)\nonumber\\
&&-E(ABC|x=1,y=1,z=1)|\leq 4,
\end{eqnarray}
%\end{widetext}
where $E(ABC|x,y,z)$'s represent the expectation value of the product of the measurement outcomes of the observables $x$, $y$, and $z$, and we call $S$ as
Svetlichny operator.
It was shown by Svetlichny \cite{svet} that quantum predictions violate his inequality, and the maximum violation ($S=4\sqrt{2}$)
allowed in quantum mechanics can be achieved with GHZ states \cite{mit}.

If we define $a=\frac{1-A}{2}$, $b=\frac{1-B}{2}$, and $c=\frac{1-C}{2}$, each of $E(ABC|x,y,z)$'s can be expressed in terms of probabilities, for example,
\begin{eqnarray}
&&E(ABC|x=0,y=0,z=0)\nonumber\\
&=&2P(a\oplus b\oplus c=xy\oplus yz\oplus xz|x=0,y=0,z=0)-1,\nonumber
\\
&&E(ABC|x=0,y=1,z=1)\nonumber\\
&=&1-2P(a\oplus b\oplus c=xy\oplus yz\oplus xz|x=0,y=1,z=1),\nonumber
\\
\end{eqnarray}
where $P(a\oplus b\oplus c=xy\oplus yz\oplus xz|x=0,y=0,z=0)$ is the probability that $a\oplus b\oplus c=xy\oplus yz\oplus xz$ under the condition $x=0,y=0,z=0$
, and $\oplus$ denotes the addition modulo $2$. So we can also write the SI as
\begin{eqnarray}
\frac{1}{8}\sum_{x,y,z}P(a\oplus b\oplus c=xy\oplus yz\oplus xz|x,y,z)\leq \frac{3}{4}.
\label{si}
\end{eqnarray}
From the above inequality we find that there is a convenient way of thinking about genuine three-particle correlations by three black boxes shared by Alice, Bob, and Carol. The correlations between inputs $x$, $y$, $z$ and outcomes $a$, $b$, $c$ are
described by probability $P(a\oplus b\oplus c=xy\oplus yz\oplus xz|x,y,z)$, and we call these boxes Svetlichny boxes \cite{barrett1}.
The maximal algebraic value $S=8$ is reached if and only
if $P(a\oplus b\oplus c=xy\oplus yz\oplus xz|x,y,z)=1$ for any $x$, $y$, and $z$.
It is obvious that Svetlichny boxes belong to tripartite NS-boxes, since Svetlichny boxes still satisfy the principle of no-signaling due to uniformly random local outcomes.

{\it Svetlichny boxes lead to violation of IC}~~Before elucidating that Svetlichny boxes can maximally violate IC, we first give a brief overview of IC.
Suppose there are two persons, Alice and Bob,
Alice has $N$ random and independent bits $(a_{1},a_{2},...,a_{N})$, and Bob receives a random variable $l\in\{1,2,...,N\}$.
Alice can send $n$ classic bits to Bob, and
Bob's task is to guess the value of the $l$-th bit in Alice's list with the help of the $n$ bits.
The amount of the information about Alice's list
gained by Bob is measured by
\begin{eqnarray}
I\equiv \sum_{k=1}^{N}I(a_{k}:g|l=k)\geq N-\sum_{k=1}^{N}h(p_{k}),
\label{i}
\end{eqnarray}
where $I(a_{k}:g|l=k)$ is Shannon mutual information between $a_{k}$ and $g$ ($g$ is Bob's guess), and $p_{k}$ is the probability
that $a_{k}=g$, both computed in the case of that Bob has received $l=k$. In Eq. (\ref{i}), the inequality can be proved by Fano inequality \cite{cover}.
IC states that physically allowed theories must have
\begin{eqnarray}
I\leq n.
\end{eqnarray}

%%%%%%%%%%%%%%%%
\begin{figure}[t]
\includegraphics[width=1\columnwidth,
height=0.6\columnwidth]{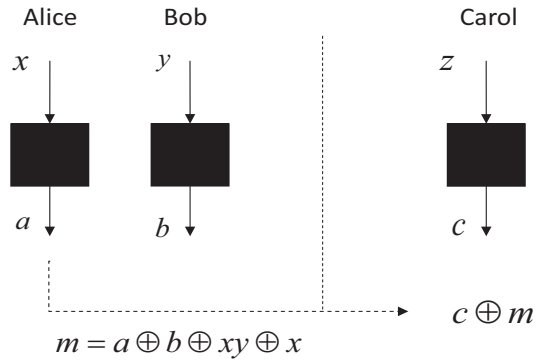} \caption{Alice, Bob, and Carol shared Svetlichny boxes ($P(a\oplus b\oplus c=xy\oplus yz\oplus xz|x,y,z)$), Alice and Bob
sit next to each other, at a long distance from Carol.
Alice or Bob sends message $m=a\oplus b\oplus xy\oplus x$ to Carol, if input $z=0$ Carol wants to learn
$x$, and if input $z=1$ she wants to learn $y$. Upon receiving message $m$ from Alice and Bob, Carol can compute her
guess $g=c\oplus m=a\oplus b\oplus c\oplus xy\oplus x$. The probabilities of correctly guessing $x$ and $y$
are Eq. (\ref{px}) and Eq. (\ref{py}). With the help of suitable Svetlichny boxes these three persons can violate IC.}
\label{fig1}
\end{figure}
%%%%%%%%%%%%%%%

Now we consider that there exist Svetlichny boxes shared by Alice, Bob, and Carol (see Fig.(\ref{fig1})). Alice and Bob
sit next to each other, at a long distance from Carol, and Alice(Bob) can send one bit to Carol. Carol's mission is to guess the
value of $x$ (Alice's input) when she receives $z=0$ and guess the value of $y$ (Bob's input) when she receives $z=1$. The message which sent by Alice (Bob) to Carol
is $m=a\oplus b\oplus xy\oplus x$. Upon receiving the message $m$, Carol can compute her
guess $g=c\oplus m=a\oplus b\oplus c\oplus xy\oplus x$. The probabilities of correct guess of $x$ and $y$ are
\begin{eqnarray}
p_{x}&=&\frac{1}{4}[P(a\oplus b\oplus c=0|0,0,0)+P(a\oplus b\oplus c=0|0,1,0)\nonumber\\
&&+P(a\oplus b\oplus c=0|1,0,0)+P(a\oplus b\oplus c=1|1,1,0)]\nonumber
\\
\label{px}
\end{eqnarray}
\begin{eqnarray}
p_{y}&=&\frac{1}{4}[P(a\oplus b\oplus c=0|0,0,1)+P(a\oplus b\oplus c=1|0,1,1)\nonumber\\
&&+P(a\oplus b\oplus c=1|1,0,1)+P(a\oplus b\oplus c=1|1,1,1)]\nonumber
\\
\label{py}
\end{eqnarray}
The Svetlichny boxes of $P(a\oplus b\oplus c=xy\oplus yz\oplus xz|x,y,z)=1$ predict $p_{x}=p_{y}=1$, from Eq. (\ref{i}) we have $I=2$ for $n=1$, so the Svetlichny boxes can maximally violate IC.

{\it The bound on genuine three-particle correlations}~~Now we proceed to show that stronger-than-quantum genuine three-particle correlations
lead to the violation of IC.

Since it is known that the maximal violation is obtained by the GHZ state \cite{seev} and in this case all probabilities $P(a\oplus b\oplus c=xy\oplus yz\oplus xz|x,y,z)$
are the same, it is a natural choice to consider the isotropic boxes and indeed this choice successfully leads to the quantum bound.
The isotropic Svetlichny boxes can be written as
\begin{eqnarray}
P(a\oplus b\oplus c=xy\oplus yz\oplus xz|x,y,z)=\frac{1+E}{2},
\label{tnsb}
\end{eqnarray}
where $0\leq E\leq1$. The Svetlichny boxes of Eq. (\ref{tnsb}) has strongest genuine tripartite correlations when $E=1$, and it correspond to uncorrelated random bits when $E=0$.
SI of Eq. (\ref{si}) is violated as soon as $E>\frac{1}{2}$, and the quantum bound $S=4\sqrt{2}$ corresponds to $E=\frac{\sqrt{2}}{2}$.

\begin{figure}[t]
\includegraphics[width=1\columnwidth,
height=0.6\columnwidth]{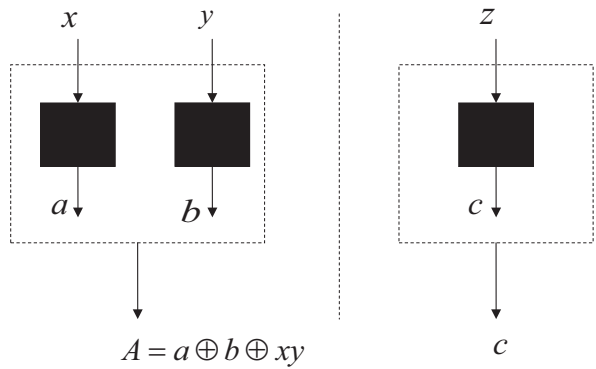} \caption{We can transform Svetlichny boxes to bipartite NS-boxes.
The left two boxes are combined to form a new box, the right box is unchanged. The only difference is that there are two input bits on the one side of this transformed NS-boxes,
while there is only one input bit on both sides of ``normal'' bipartite NS-boxes. If the initial Svetlichny boxes is described by probability
$P(a\oplus b\oplus c=xy\oplus yz\oplus xz|x,y,z)=\frac{1+E}{2}$,
the transformed bipartite NS-boxes can be described by probability $P(A\oplus c=(x\oplus y)z|x,y,z)=\frac{1+E}{2}$.}
\label{fig2}
\end{figure}

In Fig(\ref{fig2}), we illustrate how to transform
Svetlichny boxes to bipartite NS-boxes. If the initial Svetlichny boxes are described by probability $P(a\oplus b\oplus c=xy\oplus yz\oplus xz|x,y,z)=\frac{1+E}{2}$,
the transformed bipartite NS-boxes can be described by probability $P(A\oplus c=(x\oplus y)z|x,y,z)=\frac{1+E}{2}$.
So any bipartite NS-boxes of $P(a\oplus b=xy|x,y)=\frac{1+E}{2}$ can be simulated by
Svetlichny boxes of $P(a\oplus b\oplus c=xy\oplus yz\oplus xz|x,y,z)=\frac{1+E}{2}$.
In Ref. \cite{pawlowski1}, the authors proved that the bipartite NS-boxes of $P(a\oplus b=xy|x,y)=\frac{1+E}{2}$ would lead to the violation of IC
as soon as $E>\frac{\sqrt{2}}{2}$, thus we can conclude that the Svetlichny boxes of $P(a\oplus b\oplus c=xy\oplus yz\oplus xz|x,y,z)=\frac{1+E}{2}$
lead to the violation of IC as soon as $E>\frac{\sqrt{2}}{2}$. So we have proven that the maximal genuine three-particle correlation
under the constraint of IC just corresponds to the quantum bound of violation of SI.

%%%%%%%%%%%%%%%%
%%%%%%%%%%%%%%%%%%%%%%%%%%%%%%%

{\it The bound on genuine multipartite correlations}~~In Ref. \cite{seev}, the SI of three-particle has been generalized to
the case of $N$ particles. Here, by using the derivation method of Eq. (\ref{si}) we express these $N$-particle SI
in terms of probability, then genuine $N$-particle correlations can be modeled as $N$-particle no-signaling boxes (NNS-boxes).
Suppose there are $N$ players who shared $N$ particles, each one of them performs dichotomous measurements on each of the $N$ particles.
The measurement settings are represented by $x_{1}$, $x_{2}$,...$x_{N}$ respectively, with possible values $0,1$. The measurement
results are represented by $a_{1}$, $a_{2}$,...$a_{N}$ respectively, and also with possible values $0,1$. Then the $N$-particle
SI can be written as (proof in the Appendix)
\begin{eqnarray}
\frac{1}{2^{N}}\sum_{\{x_{i}\}}P\Big(\sum_{i}^{N}a_{i}=\sum_{i<j\leq N}x_{i}x_{j}|x_{1},x_{2},...,x_{N}\Big)\leq \frac{3}{4},
\label{nsi}
\end{eqnarray}
where $\{x_{i}\}$ stands for an $N$-tuple $x_{1},...,x_{N}$, $\sum_{i}^{N}$ and $\sum_{i<j\leq N}$ both denote
summation modula $2$, and $P$ is the probability that $\sum_{i}^{N}a_{i}=\sum_{i<j\leq N}x_{i}x_{j}$
with given $x_{1},x_{2},...,x_{N}$.
The isotropic NNS-boxes can be written as a
simple form:
\begin{eqnarray}
P\Big(\sum_{i}^{N}a_{i}=\sum_{i<j\leq N}x_{i}x_{j}|x_{1},x_{2},...,x_{N}\Big)=\frac{1+E}{2},
\label{nnsb}
\end{eqnarray}
where $0\leq E\leq1$. SI of Eq. (\ref{nnsb}) is violated as soon as $E>\frac{1}{2}$. The quantum bound of genuine $N$-particle correlations
corresponds to $E=\frac{\sqrt{2}}{2}$, and it can be achieved with $N$-particle GHZ states \cite{seev}.

%%%%%%%%%%%%%%%%%%%%%%%%%%%%%%%
\begin{figure}[t]
\includegraphics[width=1\columnwidth,
height=0.8\columnwidth]{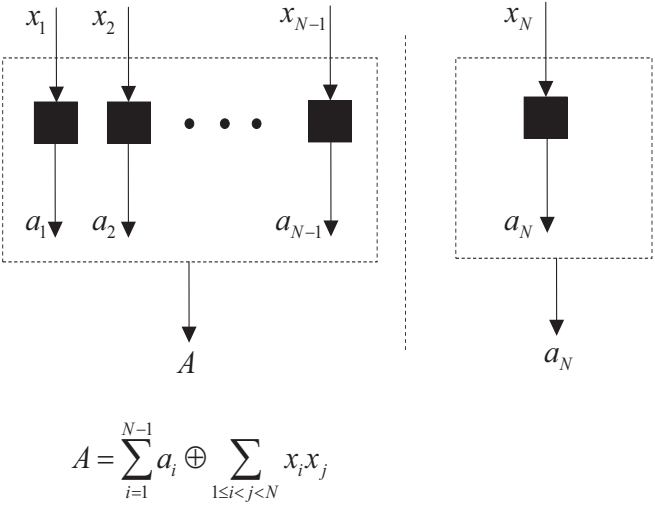} \caption{We can transform NNS-boxes to bipartite NS-boxes.
The left $N-1$ boxes are combined to form a new box, the right box is unchanged. The only difference is that there are $N-1$ input bits on the one side of this transformed NS-boxes,
while there is only one input bit on both sides of ``normal'' bipartite NS-boxes. If the initial NNS-boxes is described by probability
$P\Big(\sum_{i}^{N}a_{i}=\sum_{i<j\leq N}x_{i}x_{j}|x_{1},x_{2},...,x_{N}\Big)=\frac{1+E}{2}$,
the transformed bipartite NS-boxes can be described by probability $P(A\oplus a_{N}=(x_{1}\oplus x_{2}\oplus ...\oplus x_{N-1})x_{N})=\frac{1+E}{2}$.}
\label{fig3}
\end{figure}
%%%%%%%%%%%%%%%

In Fig.(\ref{fig3}), we illustrate the transformation of NNS-boxes to bipartite NS-boxes. If the initial NNS-boxes is
described by probability $P\Big(\sum_{i}^{N}a_{i}=\sum_{i<j\leq N}x_{i}x_{j}|x_{1},x_{2},...,x_{N}\Big)=\frac{1+E}{2}$,
the transformed bipartite NS-boxes can be described by probability $P(A\oplus a_{N}=(x_{1}\oplus x_{2}\oplus ...\oplus x_{N-1})x_{N})=\frac{1+E}{2}$.
So any bipartite NS-boxes of $P(a\oplus b=xy|x,y)=\frac{1+E}{2}$ can be simulated by
NNS-boxes of $P\Big(\sum_{i}^{N}a_{i}=\sum_{i<j\leq N}x_{i}x_{j}|x_{1},x_{2},...,x_{N}\Big)=\frac{1+E}{2}$. This implies that the NNS-boxes
of $P\Big(\sum_{i}^{N}a_{i}=\sum_{i<j\leq N}x_{i}x_{j}|x_{1},x_{2},...,x_{N}\Big)=\frac{1+E}{2}$
would lead to the violation of IC as soon as $E>\frac{\sqrt{2}}{2}$. So we have proven that the maximal genuine $N$-particle correlations
under the constraint of IC just corresponds to the quantum bound of violation of SI of $N$-particle.

{\it Discussion}
~~In this work we give an information-theoretical proof about the quantum bound of violations of
SI, i.e. the maximal violations of SI just equal to the quantum bound due to the constraint of IC.
We first employ a genuine multipartite correlation resource to simulate a bipartite correlation, and then
make use of the previously known bipartite results \cite{pawlowski1}. We note that, while there exist many different protocols to simulate a bipartite correlation by using
a genuine $N$-partite correlation, all the simulations will result in the same conclusion: if there exists a stronger-than-quantum genuine $N$-partite
correlation then we can use it to simulate a bipartite correlation which can breach IC.
With regard to different simulation protocols, for example, we can combine left $k$ boxes to form a new box and the remaining $N-k$ boxes are combined to form the other new
box. If the initial $N$-partite no-signaling boxes is described by probability $P\Big(\sum_{i}^{N}a_{i}=\sum_{i<j\leq N}x_{i}x_{j}|x_{1},x_{2},...,x_{N}\Big)=\frac{1+E}{2}$,
then the transformed bipartite no-signaling boxes is described by probability $P(A\oplus B=(x_{1}\oplus x_{2}\oplus ...\oplus x_{k})(x_{k+1}\oplus ...\oplus x_{N}))=\frac{1+E}{2}$,
where $A=\sum_{i=1}^{k}{a_{i}}\oplus\sum_{1\leq i<j\leq k}{x_{i}x_{j}}$ and $B=\sum_{i=k+1}^{N}{a_{i}}\oplus\sum_{k+1\leq i<j\leq N}{x_{i}x_{j}}$.
This simulation is different from the previous simulation but would lead to the same conclusion.

The genuine multipartite correlations are essentially more powerful correlation resources than bipartite correlations. One Svetlichny box can simulate a PR-box, but we must use
three PR-boxes to simulate a Svetlichny box \cite{barrett1}.
So the bound of genuine multipartite correlations what the IC tells us is a genuine new and exciting result, which bears some fundamental differences from
the known bipartite results.

%%%%%%%%%%%%%%%%

%%%%%%%%%%%%%%%
%\vskip 0.5 cm

%\widetext
%\section{Appendix}

{\it Appendix}~~~Proof of inequality (9)

Suppose there are $N$ players who shared $N$ particles, each one of them performs
dichotomous measurements on each of the $N$ particles.
The measurement settings are represented by $x_{1}$, $x_{2}$,...$x_{N}$, respectively, with possible values $0,1$. The measurement
results are represented by $A_{1}$, $A_{2}$,...$A_{N}$, respectively, and with possible values $-1,1$.
Then the original $N$-particle SI \cite{seev} can be expressed as
\begin{eqnarray}
S_{N}&\equiv& |\sum_{\{x_{i}\}} {v(x_{1},x_{2},...,x_{N})E(A_{1}A_{2}\cdot\cdot\cdot A_{N}|x_{1},x_{2},...,x_{N})}|\nonumber\\
&\leq& 2^{N-1},
\label{sn1}
\end{eqnarray}
where $\{x_{i}\}$ stands for an $N$-tuple $x_{1},...,x_{N}$, $E(A_{1}A_{2}\cdot\cdot\cdot A_{N}|x_{1},x_{2},...,x_{N})$ represents the
expectation value of the product of the measurement
outcomes of the observables $x_{1},x_{2},...,x_{N}$,
and $v(x_{1},x_{2},...,x_{N})$ is a sign function given by
\begin{eqnarray}
v(x_{1},x_{2},...,x_{N})=(-1)^{[\frac{k(k-1)}{2}]},
\end{eqnarray}
where $k$ is the number of times index $1$ appears in $(x_{1},x_{2},...,x_{N})$.

We can easily find that
\begin{eqnarray}
v(x_{1},x_{2},...,x_{N})=(-1)^{\sum_{i<j\leq N}x_{i}x_{j}},
\end{eqnarray}
where $\sum_{i<j\leq N}$ denotes summation modula $2$.

If we define $a_{i}=\frac{1-A_{i}}{2}$, then
\begin{eqnarray}
&&E(A_{1}A_{2}\cdot\cdot\cdot A_{N}|x_{1},x_{2},...,x_{N})=(-1)^{\sum_{i<j\leq N}x_{i}x_{j}}\nonumber\\
&\cdot&\Big[2P\Big(\sum_{i}^{N}a_{i}=\sum_{i<j\leq N}x_{i}x_{j}|x_{1},x_{2},...,x_{N}\Big)-1\Big],
\label{sn2}
\end{eqnarray}
where $\sum_{i}^{N}$ denotes summation modula $2$.
From Eq. (\ref{sn1}) and Eq. (\ref{sn2}) we finally obtain inequality (9) in the main text.

{\it Acknowledgments}
~This work is supported by National Foundation of Natural Science in
China under Grant Nos. 10947142 and 11005031.

%\newpage


\begin{thebibliography}{100}
%\bibitem{panch} S.Pancharatnam, Proc.Indian Acad. Sci. A
%{\bf44}, 247(1956).
\bibitem{bell} J. S. Bell, Physics (Long Island City, N.Y.) {\bf1},
195(1964); J. S. Bell, \emph{Speakable and Unspeakable in Quantum
Mechanics} (Cambridge University Press, Cambridge, England, 1988).
\bibitem{chsh} J. F. Clauser, M. A. Horne, A. Shimony, and R. A.
Holt, Phys. Rev. Lett. {\bf23}, 880(1969); {\bf24}, 549(E)(1970).
\bibitem{tsir} B. S. Cirel'son, Lett. Math. Phys. {\bf4}, 93(1980).
\bibitem{pr} S. Popescu and D. Rohrlich, Found. Phys. {\bf24}, 379(1994).
\bibitem{barrett1} J. Barrett, N. Linden, S. Massar, S. Pironio, S. Popescu, and D. Roberts, Phy. Rev. A {\bf71}, 022101(2005)
\bibitem{noclon} L. Masanes, A. Acin, and N. Gisin, Phys. Rev. A {\bf73}, 012112(2006);
J. Barrett, Phys. Rev. A {\bf75}, 032304(2007).
\bibitem{nobroad} H. Barnum, J. Barrett, M. Leifer, and A. Wilce, Phys. Rev. Lett. {\bf99}, 240501(2007).
\bibitem{trade} V. Scarani, N. Gisin, N. Brunner, L. Masanes, S. Pino, and A. Acin, Phys. Rev. A {\bf74}, 042339(2006).
\bibitem{key} J. Barrett, L. Hardy, and A. Kent, Phys. Rev. Lett. {\bf95}, 010503(2005);
A. Acin, N. Gisin, and L. Masanes, Phys. Rev. Lett. {\bf97}, 120405(2006).
\bibitem{dam} W. van Dam, quant-ph/0501159(2005).
\bibitem{brassard1} G. Brassard, H. Buhrman, N. Linden, A. A. M\'{e}thot, A. Tapp, and F. Unger, Phys. Rev. Lett. {\bf96}, 250401(2006).
\bibitem{nonlocal comput} N. Linden, S. Popescu, A. J. Short, and A. Winter, Phys. Rev. Lett, {\bf99}, 180502(2007).
\bibitem{brun1} N. Brunner, P. Skrzypczyk, arXiv: 0901.4070.
\bibitem{barnum} H. Barnum, S. Beigi, S. Boixo, M. B. Elliott, and S. Wehner, Phys. Rev. Lett, {\bf104}, 140401(2010).
\bibitem{acin} A. Ac\'{\i}n \textit{et al.}, Phys. Rev. Lett, {\bf104}, 140404(2010).
\bibitem{pawlowski1} M. Paw{\l}owski \textit{et al.}, Nature, {\bf461}, 1101(2009).
\bibitem{allcock} J. Allcock \textit{et al.}, Phys. Rev. A, {\bf80}, 040103(R)(2009).
\bibitem{cav} D. Cavalcanti \textit{et al.}, Nat. Comm. {\bf1}, 136(2010).
\bibitem{pironio} S. Pironio, J.-D. Bancal, and V. Scarani, J. Phys. A: Math. Theor. {\bf44}, 065303(2011).
\bibitem{svet} G. Svetlichny, Phys. Rev. D {\bf35}, 3066(1987).
\bibitem{seev} M. Seevinck and G. Svetlichny, Phys. Rev. Lett. {\bf89}, 060401(2002).
\bibitem{mit} P. Mitchell, S. Popescu, and D. Roberts, Phys. Rev. A, {\bf70}, 060101(2004).
\bibitem{cover} T. M. Cover and J. A. Thomas, \emph{Elements of Information Theory}(China Machine Press, Beijing, China, 2008).
%\bibitem{acin2} L. Masanes, A. Ac\'{\i}n, N. Gisin, Phys. Rev. A. {\bf73}, 012112(2006).
%\bibitem{cleve} R. Cleve and H. Buhrman, arXive: quant-ph/9704026.

%\bibitem{app} See Appendix

% ****** End of file template.aps ******
\end{thebibliography}
\end{document}